# Astro2020 Science White Paper

# Emission Line Mapping of the Circumgalactic Medium of Nearby Galaxies

**Thematic Areas:** Galaxy Evolution

**Principal Author:**

Name: Dennis Zaritsky
Institution: University of Arizona
Email: dennis.zaritsky@gmail.com
Phone: 520-621-6027

**Co-authors:** (names and institutions)

Peter Behroozi (University of Arizona)
Molly S. Peeples (Space Telescope Science Institute / Johns Hopkins University)
Sarah Tuttle (University of Washington)
Jessica Werk (University of Washington)
Huanian Zhang (University of Arizona)

## Abstract :

The circumgalactic medium (CGM), which harbors > 50% of all the baryons in a galaxy, is both the reservoir of gas for subsequent star formation and the depository of chemically processed gas, energy, and angular momentum from feedback. As such, the CGM obviously plays a critical role in galaxy evolution. We discuss the opportunity to image this component using recombination line emission, beginning with the early results coming from recent statistical detections of this emission to the final goal of realizing spectral-line images of the CGM in individual nearby galaxies. Such work will happen in the next decade and provide new insights on the galactic baryon cycle.

# INTRODUCTION

The key to understanding galaxy evolution is unraveling how baryons distribute themselves within dark matter halos and the processes that redistribute them over time. A full accounting of the baryons in galaxies is a long-standing but not-fully-realized goal, as evidenced by the large uncertainties remaining between the expected baryonic mass, calculated using the total mass of galaxies and the cosmological baryon fraction (Komatsu et al. 2011), and the observed baryonic mass (e.g., Miller & Bregman 2015). The required reconciliation almost certainly lies with a more complete understanding of the gaseous content within the vast dark matter halos of galaxies, but that gas is diffuse and almost certainly in multiple physical phases (e.g., Faerman et al. 2017). Various lines of investigation point to a circumgalactic medium (CGM), which harbors > 50% of all the baryons in a galaxy, that is both the reservoir of gas for subsequent star formation and the depository of chemically processed gas, energy, and angular momentum from feedback (cf. Tumlinson, Peeples, & Werk 2017). As such, the CGM plays a critical role in galaxy evolution. **We believe that in the next decade investigators will begin to map this component in individual galaxies and thereby make significant strides in an effort to address long-standing questions regarding galaxy evolution.**

Interest in tracing hydrogen gas outside the inner disks of galaxies, whether it is neutral or ionized, extends well back in time because this gas is the likely fuel reservoir for future star formation (Spitzer 1956). Deep searches for neutral gas in nearby galaxies, based on the 21cm emission feature, hit a floor column density of $\sim 10^{19}$ cm$^{-2}$ (van Gorkom 1991), resulting in apparent sharp edges to the gaseous disks of galaxies plus a limited number of isolated clouds (e.g. Minchin et al. 2003). The absence of more widely distributed H I was attributed to ionization caused by either the intergalactic UV background (Maloney 1993) or escaping radiation from the galaxy itself (Tinsley 1972; Heckman et al. 2011). This hypothesis, in turn, motivated searches for recombination radiation from ionized hydrogen in galaxy halos (e.g. Bland-Hawthorn et al. 1997). However, the emission measure of such gas is exceedingly small, demanding either deep long slit spectroscopy (Christlein et al. 2010) or integral field spectroscopy (e.g. Bland-Hawthorn et al. 1997; Dicaire et al. 2008; Hlavacek-Larrondo et al. 2011; Adams et al. 2011) and, until our recent work, yielded only upper limits at projected radii where a detection would be clearly associated with the halo gas rather than with that at the outskirts of the central galaxy.

Such emission line observations are complementary to the extensive work done using hydrogen and metal line absorption lines in QSO spectra (e.g., Young et al. 1982; Bergeron et al. 1986; Rauch 1998; Bowen et al. 2002; Werk et al. 2014). In particular, emission lines probe the local density squared rather than the column density, and, therefore, the two types of measurements together can be used to constrain the 3D density profile and the degree of clumpiness of the CGM. A key advantage of observing emission is that one can map the CGM of any galaxy, whereas absorption line studies only probe the CGM at the location of a sufficiently bright background source. The importance of observing the CGM in individual galaxies arises from the expected complexity as expressed in Figure 1.

Our theoretical understanding of accretion onto galaxies has evolved, beginning with the initial description of how infalling gas would be shock heated as it entered a dark matter halo (White & Rees 1978) to more complex treatments embedded in detailed cosmological simulations (e.g., Keres et al. 2005). A commonality of recent treatments (e.g., Nuza et al. 2014; Ford et al. 2016; Angles-Alcazar et al. 2017; Oppenheimer et al. 2018) is the resulting complexity of the halo gas and the sensitivity of important resulting details to a variety of poorly constrained factors such as the energetic feedback from the central galaxy and the intensity of the local ionizing radiation field.

In the past decade, simulations have made tremendous progress in matching observed stellar mass—halo mass relations (Naab & Ostriker 2017), but have done so with very different assumptions for stellar and black hole feedback. This has exposed a degeneracy between how much gas leaves the galaxy ("mass ejected") and how long that gas takes to return to the galaxy ("re-infall time"), which is controlled by the ejection velocity, temperature, and geometry. Increasing the mass ejected but reducing the re-infall time results in the same total number of stars formed, but gives rise to a very different structure in the CGM. Hence, theoretical progress for galaxy formation in the next decade—as well as the other science cases above—will depend on having a more complete and detailed understanding of the CGM on a galaxy-by-galaxy basis.

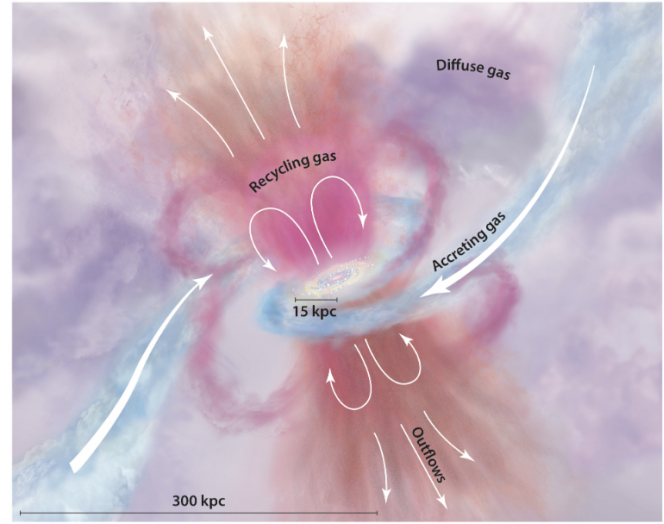

*Figure 1. This figure, from Tumlinson, Peeples, & Werk (2017), illustrates the anticipated complexity of the CGM and demonstrates why a full mapping, possible with emission line observations of the CGM, is needed to reach a deep understanding of the CGM. Ultimately, the emission and absorption line observations are complementary, but only emission line mapping will provide "images" that can be compared with simulations that test the richness of this artist's conception*

**STATISTICAL DETECTION OF EMISSION LINE RADIATION FROM THE CGM (CURRENT STATUS)**

A significant advance toward the ultimate goal of emission line mapping of the CGM in individual, nearby galaxies came from a statistical analysis of existing data (SDSS spectra from DR12; Alam et al. 2015) that improved the sensitivity to this emission by over a factor of 30 over previous studies (Zhang et al. 2016). Although stacking SDSS spectra had been done to build up absorption line detections (cf. Murga et al. 2015), it had not been done for emission studies of the CGM. The Zhang et al. (2016) study involved nearly half a million galaxies intersected by over million lines of sight and culminated in the detection of Hα and [N II] line emission to projected radii, $r_p$, of ~ 100 kpc around nearby galaxies (cf. Figure 2, taken from Zhang et al. 2018a).

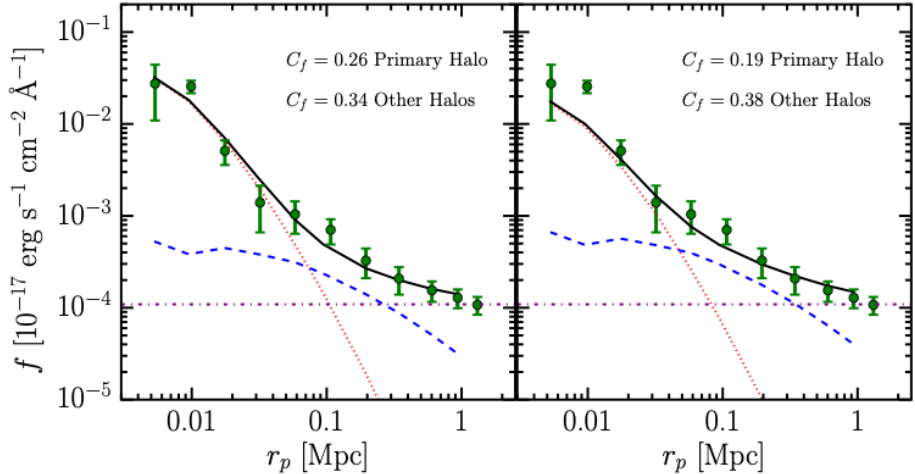

*Figure 2. Comparison of data to simple models where Zhang et al. (2018a) assign different cold gas fractions, $C_f$'s, to primary and associated halos (motivated by the differing mean masses of the two populations). All halos have smoothly distributed cold halo gas with a temperature set to 12,000 K. They show modeled Hα +[N II] emission fluxes arising from the primary halo (dotted red line), the associated halos (blue dashed line), and the systematic limit on our background measurement (horizontal line). Together (solid black line), these three components are compared to the observed emission profile (green circles with error bars). The left panel shows model results from minimization of χ2 while varying $C_f$'s. Right panel shows the model results when $C_f$'s are adopted from the relation between $C_f$ and stellar mass produced by independent hydrodynamical simulations (Ford et al. 2014). Figure from Zhang et al. (2018a).*

The detection is now at a level of detail that can be compared directly to simulations. A simple model tied to cosmological simulations in a manner described by Zhang et al. (2018a) does an excellent job at reproducing the observations (Figure 2). This agreement is maintained whether one fits the fraction of halo gas in a "cold" CGM (T=12,000 K), smooth component that follows an NFW gravitational potential profile (a fraction of 0.26 for galaxies with $M_* = 10^{10.88}$ $M_\odot$ and 0.34 for those with $M_* = 10^{10.18}$ $M_\odot$; Figure 2 left panel) or one adopts independent values of these fractions from Ford et al.'s (2014) hydrodynamical simulations, 0.19 and 0.38; Figure 2 right panel).

The results are already beginning to address some basic questions of galaxy evolution. Stacked spectra from different subsamples provide insights into the ionization state of the CGM, which shows that the CGM of high and low mass galaxies are distinctly different (Figure 3; Zhang et al. 2018b), and the role of environment, which shows the decrease in emission line flux from the CGM in galaxies when they are near a massive neighbor (Figure 4; Zhang et al. 2019).

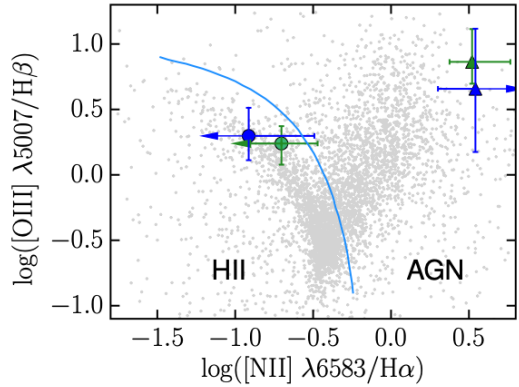

*Figure 3. Diagnostic line ratio diagram (Baldwin et al. 1981) for the CGM within $10 < (r_p/kpc) < 50$ (green symbols) and within a larger radius bin, $22 < (r_p/kpc) < 50$ (blue symbols). The circle and triangle represent the measurements for galaxies with stellar mass below and above $10^{10.4}$ $M_\odot$. The curve is the demarcation between ratios indicating ionization by star formation and AGN/shocks (Kauffmann et al., 2003). The light gray points represent the central regions of individual galaxies as measured by SDSS. Figure from Zhang et al. (2018b).*

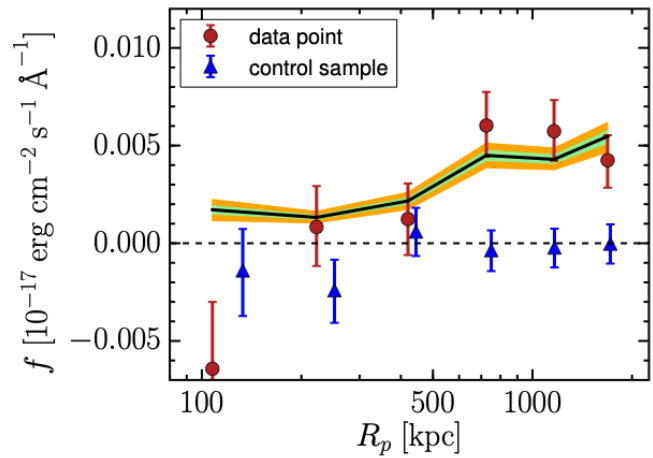

*Figure 4. A galaxy's $H\alpha +[N\,II]$ CGM emission flux $(10 < (r_p/kpc) < 50)$ vs. distance from a massive $(M^* > 10^{11}$ $M_\odot)$ galaxy (red circles). Compare to the control sample of fictitious primaries at the same distances from the same set of massive galaxy (blue triangles). The shaded region corresponds to $1\sigma$ and $2\sigma$ uncertainty of a model prediction where CGM gas is stripped in galaxies that are too close to a massive neighbor. Figure from Zhang et al. 2019.*

**WHAT WE ENVISION (THE FUTURE)**

Stacking is producing valuable constraints on the mean properties of the CGM and the relationships between those properties, the central galaxy, and its environment. However, while stacking is powerful in helping us reach extreme sensitivities with existing data, its usefulness is limited because of the lack of detail and insight it provides into specific physical processes. Ultimately, we need to map the CGM and its characteristics in individual galaxies to fully understand the range in properties within each galaxy's CGM and from one galaxy to another. **Imagine if we could make analogous measurements to those shown in Figure 2-4, but in a way that is spatially resolved within the halos of individual galaxies. If we can do that, then we can also measure the kinematics of the gas. A spatially resolved, kinematic map of the CGM in different emission lines in a range of individual galaxies is the key goal for the next decade.**

If combining millions of spectra is necessary to produce a CGM detection, how can we hope to detect the CGM in an individual galaxy? We have investigated using wide field of view spectrographs (Hectospec at the MMT and M2FS at Magellan) to provide the necessary data.

Considering the various factors involved, we estimate that in approximately several hours of observation time we can detect the ionized hydrogen of a suitable target galaxy out to 50 kpc — if we combine all of our spectra (so, unfortunately, no spatial resolution). Nevertheless, such measurements for individual galaxies, rather than for stacked galaxies, would be helpful in beginning to measure global variations in CGM properties among apparently similar galaxies.

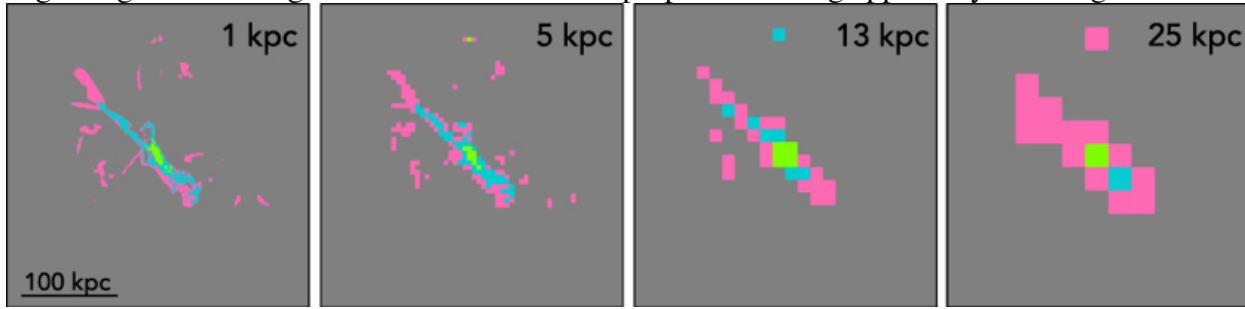

Figure 5: Here we show simulated CIII emission (z=0.2) at four different spatial resolutions (Figure 12 from Corlies & Schiminovich 2016). Even at the lowest resolution, the CGM is identifiable, while higher spatial resolution is required to provide more obvious detection of the filamentary structure.

Wide-field IFUs, such as the LLAMAS instrument being built for Magellan, will help the situation by at least an order of magnitude given their multiplexing capability. We point to the recent MUSE study of the nearby starburst ESO 338-IG04 as one such outstanding example (Bik et al. 2018). As such, we may begin being able to construct coarse emission line maps of individual galaxies (e.g., a few radial bins, or polar vs. disk plane behavior). We are not minimizing the value of such observations, which could help pin down the magnitude of the impact of feedback across a set of galaxies with different star formation or nuclear properties, but these data do not quite match our stated goal.

Emission line mapping is photon-starved and will benefit greatly from greater collecting area than that currently available. As such, the full exploitation of such maps will require the next generation of large telescope, eventually in combination with some multiplexing capability. ELTs, particularly coupled with some multiplexing capability (e.g., GMT + GMACS, with or without MANIFEST) is expected to be available within the next decade. Then we will have maps of the type shown in Figure 5 (for lines in the optical). The value of UV observations is outlined in a separate white paper (Tuttle et al.).

**SUMMARY**

We anticipate significant progress in our understanding of the baryon cycle in galaxies once we have kinematic and spatial maps of the CGM in individual galaxies. Current observations have provided measurements of the required sensitivity to emission line radiation, but detections are only available from large stacks of galaxy spectra. As such, our measurements are only of mean properties over different galaxy samples. Current state-of-the-art facilities, with existing or approved instrumentation, will provide the first coarse maps of the CGM in individual galaxies in the next 5 years. Facilities being planned for the mid to late 2020's, will provide the first resolved kinematic and spatial emission line maps, providing measurements of the flows of ionized gas throughout galaxy halos.